\documentclass{PoS}
\usepackage{braket}
\usepackage{graphicx}
\usepackage{amsmath,amssymb}
\usepackage{fancyhdr}

\title{Study of intermediate states in the inclusive semileptonic $B \rightarrow X_c l \nu$ decay structure function}

\ShortTitle{Study of intermediate states in the inclusive semileptonic $B \rightarrow X_c l \nu$ decay structure function}
\author{JLQCD Collaboration: \speaker{G. Bailas}$^{a}$\thanks{E-mail: gabriela@kek.jp}, S. Hashimoto$^{a,b}$, T. Kaneko$^{a,b}$, J. Koponen$^{a}$\\
        {$^{a}$ High Energy Accelerator Research Organization (KEK), Ibaraki 305-0801, Japan} \\
        $^{b}$  School of High Energy Accelerator Science, SOKENDAI (The Graduate University for Advanced Studies), Ibaraki, 305-0801, Japan\\
   }

\abstract{We analyze the inclusive semileptonic $B \to X_c \ell\nu$ structure functions in 2+1-flavor lattice QCD. The M\"obius domain-wall fermion action is used for light, strange, 
charm and bottom quarks. The structure function receives contributions from various exclusive modes, including the dominant S-wave states $D^{(*)}_s$ as well as the P-wave states $D_s^{**}$. We can identify them in the lattice data, from which we put some constraints on the $B_s \to D_s^{**}\ell\nu$ form factors.
}

\FullConference{37th International Symposium on Lattice Field Theory - Lattice2019\\
		16-22 June 2019\\
		Wuhan, China}

\begin{document}

\section{Introduction}
The semileptonic decays of $B$ meson to excited states of $D$ meson have not been well understood. In this work, we focus on the decays $B\to D^{**}\ell\nu$, where $D^{**}$ stands for one of orbitally excited states $D_0^*$, $D_1^\prime$, $D_1$ and $D_2^*$. Among them, the former two have large width of order $200-400$~MeV, while the other two are narrower, $30 - 50$~MeV \cite{Tanabashi:2018oca}. There are some theoretical estimates on these semileptonic decay rates, which suggest that the narrower states have much larger rates than the broader ones \cite{Uraltsev:2000ce}. The experimental data do not support this expectation and the problem remains for more than a decade. In this study, we use lattice QCD calculation to obtain some insight into this problem. 

In general, excited states are more difficult to treat in lattice calculations, because of larger statistical noise. Preparing an interpolating operator that efficiently creates the desired state is also challenging since the states have non-trivial structures. Here, we use the forward scattering matrix element of $B$ meson, which was developed for a calculation of inclusive decay structure function \cite{Hashimoto:2017wqo}. It does not require an explicit identification of the individual states. Rather, the states are created through a flavor changing vector or axial-vector current just as in the physical process of $B\to D^{**}\ell\nu$. From a relevant four-point function, we are able to identify the corresponding contributions. 

The rest of the paper is organized as follows. Section \ref{section2} and \ref{section2.1} introduce the form factors for the P-wave states. Section \ref{section3} presents our lattice computation strategy. Section \ref{section4} contains our results and conclusions for the zero and non-zero recoil cases. Finally, Section \ref{section6} presents our conclusions.

\section{P-wave states $D^{**}$ and their form factors}\label{section2}
In the static limit $(m_b, m_c \to \infty)$, the heavy quark symmetry emerges and the meson spectrum can be constructed by combining the spin $1/2$ of the heavy quark with the total angular momentum and parity $j^{P}$ of the light degrees of freedom (light quarks and gluons). For instance the S-wave states of total spin-parity $0^{-}$ and $1^{-}$ become degenerate. For the P-wave states, the light degrees of freedom may have $j^{P} = (1/2)^{+}$ or $(3/2)^{+}$, which combined with the heavy quark spin produce the states of $J^{P} = (0^{+}, 1^{+})$ as well as $(1^{+}, 2^{+})$, respectively. When $m_b$ and $m_c$ are finite the states are classified according to their parity $P$ and total angular momentum $J$. They are named $D^{*}_0$, $D_1^{'}$, $D_1$ and $D_2^{*}$, respectively.  

In the heavy quark limit the relevant matrix elements for $B \to D^{**} \ell\nu$ decays can be parametrized by two form factors, the Isgur-Wise functions $\tau_{1/2}$ and $\tau_{3/2}$ \cite{Isgur:1991wr}:
\begin{flalign}\label{isgur1} \nonumber
&\braket{D_0^{1/2}(v')|\bar{c}\gamma_5\gamma_{\mu}b|B(v)} \propto \tau_{1/2}(w)(v-v')_{\mu}, \\ 
&\braket{D_2^{3/2}(v', \epsilon)|\bar{c}\gamma_5\gamma_{\mu}b|B(v)} \propto \tau_{3/2}(w)\left((w+1) \epsilon^{*}_{\mu\alpha}v^{\alpha} - \epsilon^{*}_{\alpha\beta} v^{\alpha}v^{\beta} v^{\prime}_{\nu}  \right),
\end{flalign}
where $v$ and $v'$ are the velocities associated with the $B$ and $D^{**}$ mesons respectively, $w = (v' \cdot v)$ and $\epsilon$ is the polarization tensor of the $D^{**}$ mesons. Previous theoretical estimates were obtained through sum rules. The most relevant one in this context was derived by Uraltsev \cite{Uraltsev:2000ce}, 
\begin{eqnarray}\label{ulratsev}
\sum_{n} \left( |\tau_{3/2}^{(n)}(1)|^2 - |\tau_{1/2}^{(n)}(1)|^2 \right) = \frac{1}{4},
\end{eqnarray}
where $\tau^{(0)}_{1/2} \equiv {\tau}_{1/2}$, $\tau^{(0)}_{3/2} \equiv {\tau}_{3/2}$ and the sum over $n$ is done for all $j =$ $1/2$ and $3/2$ states. One may expect saturation from the ground states, leading to $ |\tau_{3/2}^{(0)}(1)|^2 - |\tau_{1/2}^{(0)}(1)|^2 \approx \frac{1}{4} $ and consequently to $|\tau_{1/2}(1)| < |\tau_{3/2}(1)|$. The sum rule concerns the zero-recoil limit $(w = 1)$, where the $B$ and the $D^{**}$ mesons have the same velocity. To obtain the decay rates, however, one has to integrate over $w$, and one assumes that the inequality remains for $w \neq 1$.

BaBar \cite{Aubert:2008ea} and BELLE \cite{Liventsev:2007rb} have measured the composition of the semileptonic $B \to X_c l \nu$ decay. It turned out that $X_c$ is $70\%$ composed by $D$ and $D^{*}$ mesons (S-wave states) and $15\%$ by $D_{1}$ and $D_2$ ($j = 3/2$ states). A natural candidate for the remaining $15\%$ would be the $D_{0}$ and $D^{\prime}_1$ ($j = 1/2$ states) mesons. However, this proposal seems to be in conflict with the theoretical prediction discussed above, which has been called the ``$1/2$ versus $3/2$ puzzle''.

\section{Form factors of $B \to D^{**}l\nu$}\label{section2.1}
The Isgur-Wise form factors (\ref{isgur1}) are not used directly in our work. We use the conventional definition of the $B \to D^{**}$ form factors given by

\begin{itemize}
\item P-wave $j^{+} = \frac{3}{2}^{+}$ states:
\begin{flalign}\nonumber\label{P_1}
&\sqrt{M_BM_{D_1}}^{-1}\braket{D_1(v', \epsilon)|V_{\mu}|B(v)} = f_{V_1}\epsilon^{*}_{\mu}+(f_{V_2}v_{\mu}+f_{V_3}v'_{\mu})(\epsilon \cdot v),\\
&\sqrt{M_BM_{D_1}}^{-1}\braket{D_1(v', \epsilon)|A_{\mu}|B(v)} = - if_A\epsilon_{\mu\alpha\beta\gamma}\epsilon^{*}_{\alpha}v_{\beta}v'_{\gamma},
\end{flalign}

\item P-wave $j^{+} = \frac{1}{2}^{+}$ states:
\begin{flalign}\nonumber \label{P_2}
&\braket{D_0^{*}(v')|V_{\mu}|B(v)} = 0, \\ \nonumber
&\sqrt{M_BM_{D^*_0}}^{-1}\braket{D_0^{*}(v')|A_{\mu}|B(v)} = g_{+}(v_{\mu} +v'_{\mu}) + g_{-}(v_{\mu} - v'_{\mu}), \\ \nonumber
&\sqrt{M_BM_{D^*_1}}^{-1}\braket{D_1^{*}(v', \epsilon)|V_{\mu}|B(v)} = g_{V_1}\epsilon^{*}_{\mu} + (g_{V_2}v_{\mu} + g_{V_3}v'_{\mu})(\epsilon^{*}\cdot v),\\ 
&\sqrt{M_BM_{D^*_1}}^{-1}\braket{D_1^{*}(v', \epsilon)|A_{\mu}|B(v)} = ig_A \epsilon_{\mu \alpha \beta \gamma}\epsilon^{*}_{\alpha}v_{\beta}v'_{\gamma}.
\end{flalign}
\end{itemize}
Here, $f_{V_1}$, $f_{V_2}$, $f_{V_3}$ and $f_A$ are form factors for $j^{+} = {3}/{2}^{+}$ and $g_+$, $g_-$, $g_{V_1}$, $g_{V_2}$, $g_{V_3}$ and $g_A$ represent the ones for $j^{+} = {1}/{2}^{+}$. They are functions of $w = v \cdot v'$.

In the heavy quark expansion, these form factors can be written in terms of the Isgur-Wise functions $\tau_{3/2}(w)$, $\tau_{1/2}(w)$ plus the terms to represent the $1/m_c$ and $1/m_b$ corrections. Such calculation was performed for the P-wave decay modes \cite{Leibovich:1997em}, which we use in the following.

\section{Lattice computation strategy}\label{section3}

We utilize the forward-scattering matrix element, which represents the inclusive decay of the $B$ meson. It contains contributions from all possible final states with a certain weight factor. We compute a four-point function corresponding to the matrix element:
\begin{eqnarray}\label{4point}
C_{\mu\nu}^{JJ}(t;\vec{q}) = \int d^3\vec{x}e^{i\vec{q}\cdot {\vec{x}}}\frac{1}{2M_B}\braket{B(\vec{0})|J_{\mu}^{\dagger}(\vec{x},t)J_{\nu}(0)|B(\vec{0})},
\end{eqnarray}

\begin{figure}[h]
\centering
\includegraphics[width=23pc]{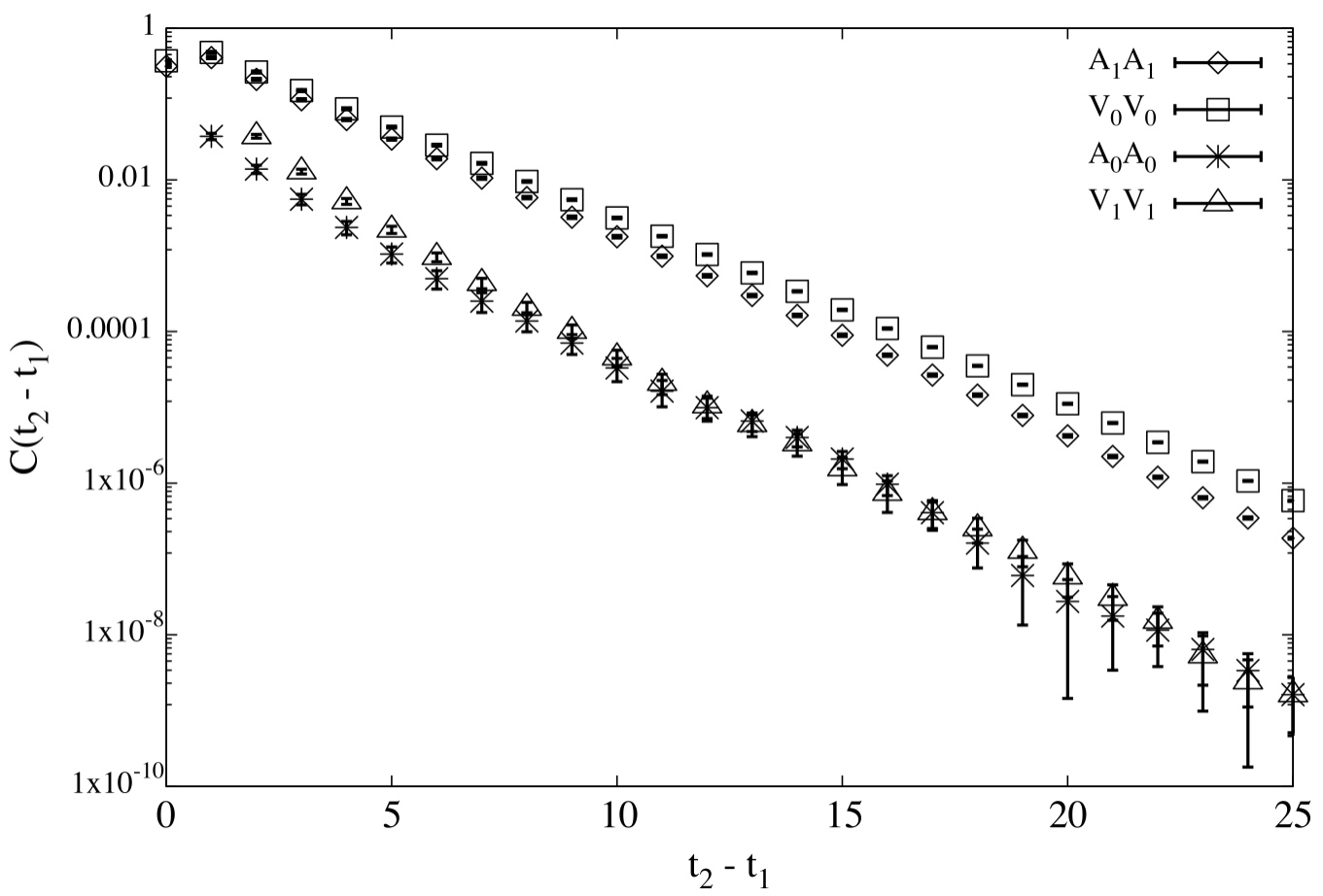}\hspace{2pc}%
\caption{\label{results_1} Four-point correlator of double insertions of vector and axial currents.}
\end{figure}

which can be extracted by taking a ratio of the four-point function to two-point functions as
\begin{eqnarray}\label{ratios}
\frac{C_{\mu\nu}^{SJJS} (t_{snk}, t_1, t_2, t_{src})}{C^{SL}(t_{snk}, t_2)C^{LS}(t_1, t_{src})} \to \frac{\frac{1}{2M_B}\braket{B(\vec{0})|J_{\mu}(\vec{q}, t_1)^{\dagger}J_{\nu}(\vec{q}, t_2) |B(\vec{0})}}{ \frac{1}{2M_B}|\braket{0|P^L|B(\vec{0})}|^2              } .
\end{eqnarray}
For details about other uses of (\ref{4point}) we refer to \cite{Hashimoto:2017wqo}. An example of the correlator ratio (\ref{ratios}) is shown in Fig. \ref{results_1}. It is plotted as a function of $t_2 - t_1$, the separation between two current insertions. One can see the exponential fall-off due to charm quark propagation. 

In the zero-recoil limit, $\vec{q} = \vec{0}$, there are two distinct channels. One is those for the temporal vector current $V^0$ and the spatial axial-vector current $A^k$, which are the upper two lines in Fig. \ref{results_1} and correspond to the S-wave final states $D$ and $D^{*}$, while for $A^0$ and $V^k$ the final states have an opposite parity and they correspond to the P-wave states shown by the lower two lines. Therefore, for sufficiently large separations $t_2 - t_1$, the final states are dominated by the $D^{**}$ states and the exponential fall-off may be written in terms of the corresponding decay form factors:
\begin{flalign}\nonumber \label{fithere}
&C^{A_0A_0}(t) = |g_{+}(1)|^2e^{-m_{D^{*}_{0}}t},\\
&C^{V^kV^k}(t) = \frac{|g_{V_1}(1)|^2}{4}e^{-m_{D^{*}_1}t} + \frac{|{f_{V_1}}(1)|^2}{4}e^{-m_{D_1}t}.
\end{flalign}
Then, we are able to extract the form factor $|g_{+}(1)|$ as well as a combination of $|g_{V_1}(1)|$ and $|{f_{V_1}}(1)|$ by fitting the lattice data at large time separations $t = t_2 - t_1$. The contribution from $|g_{V_1}(1)|$ is small for a reason that we describe later. 

\section{Results}\label{section4}

We have performed a set of lattice QCD simulations with $2+1$ flavors of dynamical quarks using the tree-level improved Symanzik gauge action and the M\"obius domain-wall fermions. Our computation preserves chiral symmetry and has a large lattice cutoff $a^ {-1} \simeq 2.5 - 4.5$ GeV. The strange quark mass $m_s$ is simulated close to its physical value, whereas the degenerate up and down quark mass $m_{ud}$ corresponds to pion masses as low as $M_{\pi} \sim 230$ MeV. In this work, we use a $48^3 \times 96$ lattice of $a = 0.055$ fm. The spatial lattice size $L$ satisfies the condition $M_{\pi} L \gtrsim 4 $ to control finite volume effects. The charm quark mass $m_c$ is set to its physical value, whereas we take bottom quark mass $m_b = 1.25^{4}m_c$, which is smaller than the physical value. The spectator quark in this calculation is strange, so that the relevant decays are actually those of $B_s \to D^{**}_{s} l \nu$. More information about our lattice data can be found in \cite{Nakayama:2016atf}. The statistics is $100$ independent gauge configurations with $4$ source locations on each configuration.

The form factors for P-wave final states introduced in (\ref{P_1}) and (\ref{P_2}) can be expanded in $1/m_c$ and $1/m_b$ as~\cite{Leibovich:1997em}
\begin{flalign} \label{uma}
&\sqrt{6}f_{V_1}(w) = - [w^2 - 1 + 8\epsilon_c (\bar{\Lambda}' - \bar{\Lambda})] \tau(w) + ..., \\
&g_{+}(w) = - \frac{3}{2} (\epsilon_c + \epsilon_b)(\bar{\Lambda}^{*} - \bar{\Lambda})\zeta(w) + ..., \\
\label{duas}
&g_{V_1}(w) = [w-1+(\epsilon_c - 3\epsilon_b)(\bar{\Lambda}^{*} - \bar{\Lambda})]\zeta(w)+...,
\end{flalign}
where $\zeta(w) = 2\tau_{1/2}(w)$ and $\tau(w) = \sqrt{3}\tau_{3/2}(w)$ are the Isgur-Wise functions, $(\bar{\Lambda}^\prime - \bar{\Lambda})$ and  $(\bar{\Lambda}^* - \bar{\Lambda})$ stand for the mass difference between S-wave and P-wave states and $\epsilon_c = 1/2m_c$, $\epsilon_b = 1/2m_b$. As anticipated, the form factors for the P-wave states vanish in the zero-recoil limit when $\epsilon_b  = \epsilon_c = 0$, because of the parity conservation. Away from the heavy quark limit, small contribution arises at the order of $\epsilon_c$ and $\epsilon_b$, which explain the small amplitudes of two lower lines found in Fig. \ref{results_1}. 

Using $(\bar{\Lambda}^* - \bar{\Lambda}) = 0.36$ MeV and $(\bar{\Lambda}^\prime - \bar{\Lambda}) = 0.40$ MeV following \cite{Bernlochner:2016bci}, we extract $\tau(1)$ and $\zeta(1)$ from the lattice data of (\ref{fithere}). The contribution of $|g_{V_1}(1)|$ is neglected because $\epsilon_c - 3\epsilon_b$ in (\ref{duas}) is numerically small. Our results are $\tau_{3/2}(1) = 0.45(7)$ and $\tau_{1/2}(1) = 0.39(6)$. This suggests that $\tau_{3/2}(1) \sim \tau_{1/2}(1)$, which is in agreement with the experimental results $\Gamma(B \to D^{**}_{1/2}l \nu) \approx \Gamma (B \to D^{**}_{3/2} l \nu)$. Our result is also consistent with the phenomenological analysis of the experimental data \cite{Bernlochner:2016bci}. A previous lattice calculation was done in the heavy quark limit \cite{Blossier:2009vy}. Its results $\tau_{3/2}(1)~=~0.53(2)$ and $\tau_{1/2}(1) = 0.30(3)$ favor the sum rule expectation $\tau_{3/2}(1) > \tau_{1/2}(1)$. Our result is not inconsistent with theirs within a large error. However, statistical error has to be reduced before drawing any firm conclusions. 

\begin{figure}[h]
\centering
\includegraphics[width=23pc]{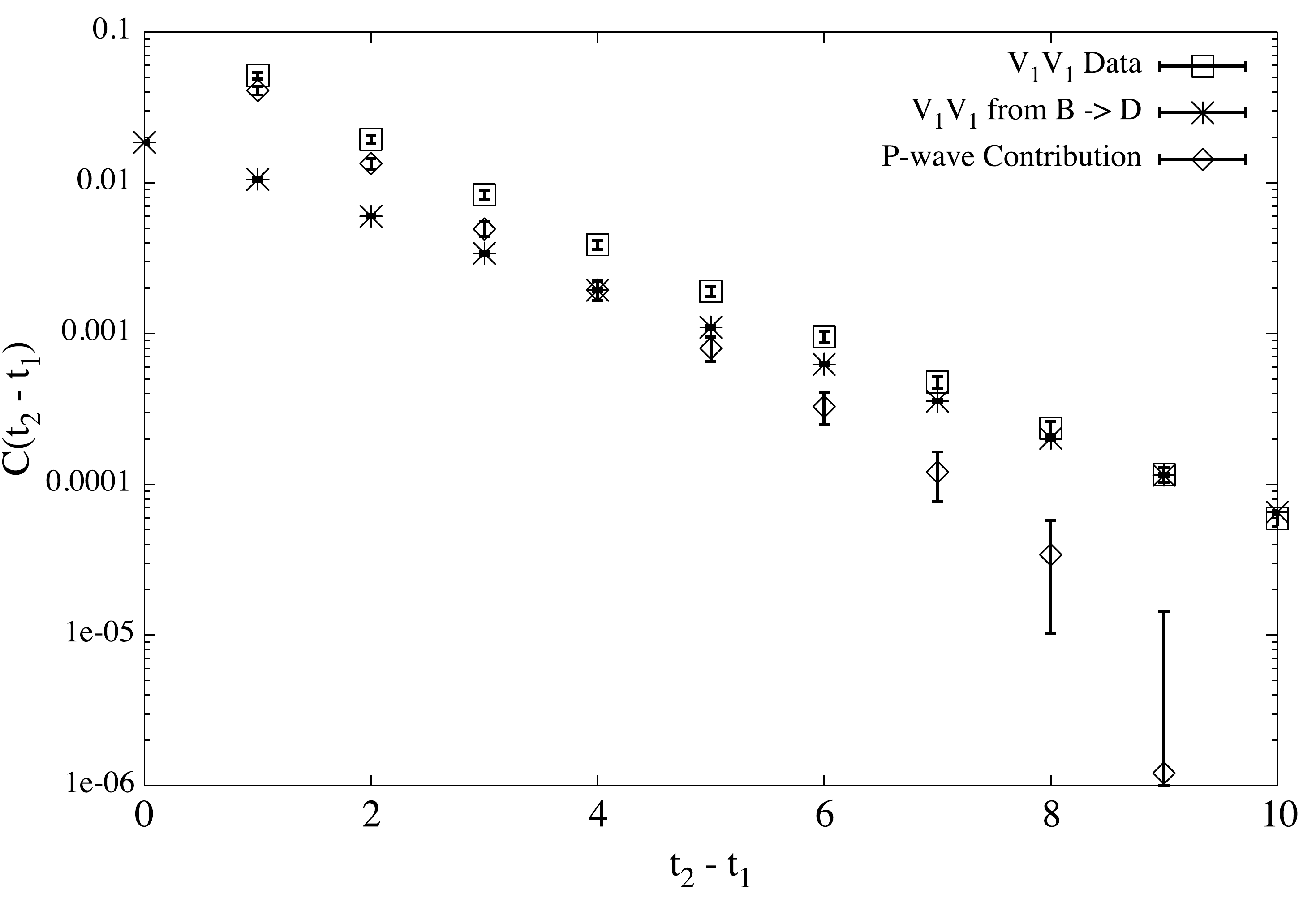}\hspace{2pc}%
\includegraphics[width=23pc]{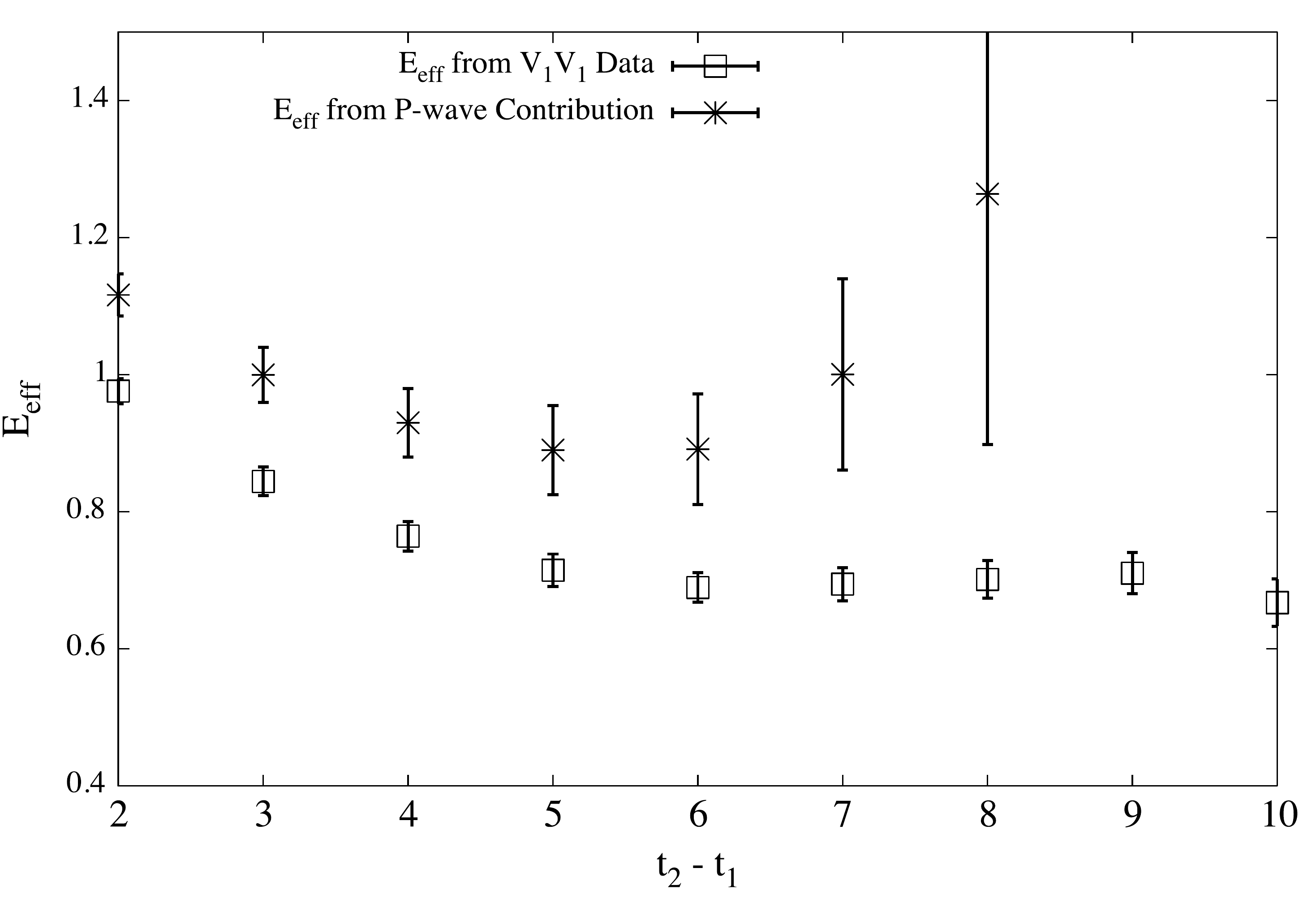}\hspace{2pc}%
\caption{\label{V1V1_001_wsub} On the left-side we have the four-point functions ratios for the different lattice data. On the right-side we present the effective energy result for the P-wave states.}
\end{figure}

Inserting finite momentum in the final state, both S-wave and P-wave states contribute to (\ref{4point}). Since the P-wave amplitude is relatively small, we need to carefully subtract the S-wave states to extract the P-wave contributions. In Fig. \ref{V1V1_001_wsub} (upper plot) the curve entitled ``$V_1V_1$ from $B \to D$'' represents the expected S-wave states, which is a reconstructed from the lattice calculation dedicated for the $B \to D^{(*)} l \nu$ decay \cite{Kaneko:2018mcr}. The curve named ``P-wave Contributions'' is obtained after subtracting the S-wave contribution. One can clearly find that the lower energy state corresponding to the S-wave contribution is removed and the higher energy state is left (see also the effective mass plot Fig. \ref{V1V1_001_wsub} (bottom) before (square) and after the subtraction (star)). 

With a momentum insertion in the Z direction, i.e. momentum $p' = \frac{2\pi}{L} (0,0,1)$, we extract the form factors $f_{V_1}$ and $f_{V_3}$ from perpendicular $V_1V_1$ and longitudinal $V_3V_3$ vector channels, res-pectively. Following the 
``Approximation~$A$'' of \cite{Leibovich:1997em, Bernlochner:2016bci}, which means ${O}(w-1) \sim {O}(\epsilon_{c,b})$ and their higher orders are truncated, we use
\begin{flalign} \label{approxA}
&f_{V_1}(w) = \frac{1}{\sqrt{6}}[(1-w^2)\tau(w) - 4\epsilon_c(w+1)(w\bar{\Lambda}' - \bar{\Lambda})\tau(w)             ]       
\end{flalign} 
and
\begin{flalign} \label{approxA1}
&wf_{V_1}(w) + (w^2 - 1)f_{V_3}(w) = \frac{w}{\sqrt{6}}[(1-w^2)\tau(w) - 4\epsilon_c(w+1)(w\bar{\Lambda}' - \bar{\Lambda})] + \frac{w^2 - 1}{\sqrt{6}}(w-2)\tau(w)
\end{flalign} 
to extract the Isgur-Wise form factor $\tau(w)$. In the same way, we also obtain $\zeta(w)$ from the form factor $g_{+}$ through the $A_0A_0$ channel. Our results are $\tau(w)=0.539(33)$ for $V_1V_1$, $\tau(w) = 0.455(27)$ for $V_3V_3$ and $\zeta(w) = 1.21(14)$, at $w = 1.027$. The inconsistency between $V_1V_1$ and $V_3V_3$ may be due to the approximation involved in the analysis.

The slopes of the Isgur-Wise functions $\tau'(w)$ and  $\zeta'(w)$ defined through
\begin{eqnarray}
\tau(w) = \tau(1)[1+\tau'(w-1)],\\
\zeta(w) = \zeta(1)[1+\zeta'(w-1)],
\end{eqnarray} 
are obtained combining with the zero-recoil results. We obtain $\tau'(1) = -7.8(6.3)$ for $V_1V_1$, $\tau'(1) = -12.2(5.3)$ for $V_3V_3$ and $\zeta'(1) = 21(12)$. Even with the large error, our final results are in agreement with the phenomenological results \cite{Bernlochner:2016bci}.

\section{Discussions}\label{section6}

The results shown in this write-up are a by-product of a calculation of the inclusive
decay structure functions \cite{Hashimoto:2017wqo}. By inspecting the energy and amplitude
of the final states contributing to the forward-scattering matrix elements, we are able to
identify those states as the P-wave D mesons, which is natural since they are contributing 
to the physical processes as experimentally observed. The method to extract the excited state
contribution is not particularly superior compared to dedicated calculations because the 
statistical noise is larger for four-point functions. It may be useful however, when a proper
interpolating operator is not known for the excited states like those for the $j = 1/2$ states.
Also, this work provides a good consistency test of the strategy to obtain the inclusive 
decay structure functions.

\subsection*{Acknowledgment}

Numerical computations are performed on Oakforest-PACS at JCAHPC.
This work was supported in part by JSPS KAKENHI Grant Number JP18H03710 
and by MEXT as "Priority Issue on post-K computer".


\end{document}